\documentclass{article}
\usepackage{hyperref}

\usepackage{arxiv}

\usepackage[utf8]{inputenc} 
\usepackage[T1]{fontenc}    
\usepackage{hyperref}       
\usepackage{url}            
\usepackage{booktabs}       
\usepackage{amsfonts}       
\usepackage{nicefrac}       
\usepackage{microtype}      
\usepackage{lipsum}
\usepackage{graphicx}
\graphicspath{ {./images/} }

\title{AirCalypse: Can Twitter Help in Urban Air Quality Measurement and Who are the Influential Users?}

\author{
 Prithviraj Pramanik\thanks{ 
Corresponding Author: \href{mailto:prithvirajpramanik@yahoo.co.in}{prithvirajpramanik@yahoo.co.in}; 
\textbf{Conference Reference}
This paper was published in \textit{WWW '20: Companion Proceedings of the Web Conference 2020},  pp. 450-445, 2020. DOI: \href{https://doi.org/10.1145/3366424.3382120}{10.1145/3366424.3382120}.
 }\\
  Department of CSE\\ 
  NIT Durgapur\\
  India-713209\\
   \And
 Tamal Mondal\\
  Department of CSE\\ 
  NIT Durgapur\\
  India-713209\\
  \And
Subrata Nandi\\
  Department of CSE\\ 
  NIT Durgapur\\
  India-713209\\
   \And
  Mousumi Saha\\
  Department of CSE\\ 
  NIT Durgapur\\
  India-713209\\
}

\hypersetup{
    pdfinfo={
        Title={AirCalypse: Can Twitter Help in Urban Air Quality Measurement and Who are the Influential Users?},
        Author={Prithviraj Pramanik, Tamal Mondal, Subrata Nandi, Mousumi Saha},
        Subject={Published in [WWW '20: Companion Proceedings of the Web Conference 2020]},
        Keywords={Social Media, Participatory Sensing, Air Quality, Influential User},
        DOI={10.1145/3366424.3382120}
    }
}

\begin{document}
\date{}

\maketitle


\begin{abstract}
In this digital age, Online Social Media's ubiquity has led it to it's role as a "Sensor". Starting from disaster response to political predictions, online social media like Twitter, have been instrumental and are actively researched areas. In this work, we have focused on something quite insidious in the current context, i.e., air pollution in developing regions. Starting as an empirical study on using Twitter as a "Sensor" to measure air quality, the focal point of this work is to identify the users who have been actively tweeting in the air pollution events in Delhi, the capital of India. From these users, we try to identify the influential ones, who play a significant role in creating the initial awareness and hence act as "Sensors". We have utilized a tailored "TRank" algorithm for finding out the influential users by considering \textit{Retweet, Favorite, and Follower influence} of the users. After ranking the users based on their social influence, we further study the behavior, i.e., perception of pollution from those users' posts with respect to the actual air pollution levels using the physical sensors. The tracking of influential users in air quality monitoring assists in developing a crowd sensed air quality measurement framework, which can augment the physical air quality sensors for raising awareness against air pollution.               
\end{abstract}


\section{Introduction} \label{intro}

\begin{quote}
Ambient (outdoor) air pollution in both cities and rural areas was estimated to cause 4.2 million premature deaths worldwide per year in 2016; this mortality is due to exposure to small particulate matter of 2.5 microns or less in diameter (PM2.5), which cause cardiovascular and respiratory disease, and cancers.\\\null \hfill - World Health Organization \footnote{https://www.who.int/en/news-room/fact-sheets/detail/ambient-(outdoor)-air-quality-and-health, May 2018} 
\end{quote}

Increasing air pollution has become a major threat to the well being of human health across the globe irrespective of the economic status of the countries. The insidious nature of air quality further checks the awareness required to take the necessary precautions against it. Especially in a developing country like India, such an increment in air pollution is responsible for numerous health hazards. It has created a fertile ground for diseases related to heart, lung, eyes, among other organs, which leads to premature deaths \cite{guttikunda2014nature, gupta2008valuation}. Hence, there is a requirement of a strong air quality monitoring system to help the administration to bring about changes in the policies to improve air quality.  According to the report of Institute for Health Metrics and Evaluation (IHME) \footnote{http://www.healthdata.org/} of $2013$ and Global Burden of Disease (GBD) Assessment $2017$ \cite{balakrishnan2019impact}, India is one of the leading nations where the increasing air pollution level is a major health risk factor, causing millions of unnatural deaths and health hazards in several urban regions like Delhi, Kanpur, Kolkata, Kochi, among others and suburban regions like Korba (Chattisgarh), Gaziabad (Outside of Delhi) among others. In major cities in India, the pollutants which primarily originate from air pollution are Particulate Matters (PM) (Mainly $PM_{10}, PM_{2.5}$). From studies \cite{gupta2008valuation,guttikunda2014nature}, it is evident that the air pollution levels expressed in the concentration of the \textit{PM} at metro cities like Delhi, Kanpur among others exceed the World Health Organisation (WHO)\footnote{https://www.who.int/phe/health\_topics/outdoorair/databases/AAP\_database\_\\summary\_results\_2016\_v02.pdf, Last Accessed: 15th January, 2020} and the National Ambient Air Quality Standards (NAAQS) standard \cite{Chowdhury10711} for a majority part of the year. Furthermore, it has also been reported that such increased contamination affects human lives through escalating premature fatalities in the future \cite{balakrishnan2019impact}, and there lies the insidious nature of air pollution. Hence, awareness of air quality is critical for long term well-being.

For evaluating the present status of air quality, measuring and regulating pollution levels according to WHO and NAAQS standards, the Central Pollution Control Board of India (CPCB) has constructed several Continuous Ambient Air Quality Monitoring Stations (CAAQMS) across different cities in India including Delhi. However, the coverage area of such AQM stations is limited to $5\%$ of the total country according to the yardstick of CPCB's estimate for India \footnote{http://www.urbanemissions.info/blog-pieces/air-monitoring-101/, Last Accessed: 15th January 2020}. Only in capital city "Delhi", there are $17$ CAAQMS stations\footnote{https://app.cpcbccr.com/ccr\_docs/caaqms\_list\_NCR.pdf, Last Accessed: 15th January 2020} to cover the area of $1484$ sq. km out of the $207$ CAAQMS\footnote{https://app.cpcbccr.com/caaqms/download?filename=Station\_List.xlsx, Last Accessed: 15th January} deployed to cover all of India, which, has a total area of 3.3 million sq. Km \footnote{https://www.india.gov.in/india-glance/profile, Last Accessed: 15th January, 2020}.

Naturally, due to the sparse deployment of CAAQM stations, gathering air quality information cannot always be accomplished with the current CAAQM station count. Furthermore, the installation and annual maintenance costs for each of such CAAQM station is also huge \cite{pramanik2018aircalypse}. \\
In the past few years, it is observed that the communities of urban regions have started creating their perceptions online through Online Social Media like Twitter, Sina-Weibo regarding a variety of topics. This has helped in prediction of the events like election results \cite{bermingham2011using},  detection of influenza \cite{signorini2011use} and gathering situational information from disaster events \cite{vieweg2010} with high veracity and velocity. 
From our prior work \cite{pramanik2018aircalypse}, we have started to try and understand if such predictions can be used for detecting this growing air pollution and find out what the different signals and noise in the data stream are. Primarily, the users are trying to express their reactions, sentiments, and opinions towards pollution level through:
\begin{enumerate}

\item Portraying their concern, criticism, and severe discomfort.
\item Providing insights on the affecting pollutants and their various sources. 
\item Suggesting actions and precautions which should be considered by the concerned authority and citizens of urban regions. 
\end{enumerate}
India, which has been rapidly acquiring a huge Twitter base over the years, provides the perfect place to understand such perceptions show that there are definite patterns in pollution tweets. In fact, in the recent past ($2017 - 2018$), there were several occasions where the pollution-related tweets have been trended countrywide on Twitter due to the rise in air pollution levels in New Delhi, India \cite{pramanik2018aircalypse}. 

Now, the question lies in whether such community perceptions, i.e., reactions can be utilized along with the CAAQM data for developing a better air quality monitoring system with high accuracy? Whether the Twitter-specific aspects can be used for mapping of community reactions with the pollution level? In the state of the art \cite{jiang2015using,pramanik2018aircalypse}, limited work has been performed in this regard. \textit{Jiang et al.} in \cite{jiang2015using} analyzed the correlation between user-related posts in "Sina-Weibo" regarding pollution level and daily Air Quality Index values in Beijing, China. They have established the association between social perception and AQI to some extent. However, in India, like the rest of the world, Twitter is the major micro-blogging platform and is distinctively different from Sina-Weibo \cite{gao2012comparative}. Besides, in our previous work \cite{pramanik2018aircalypse}, we tried to explore the latent perception between CAAQM data and tweet trends of the community regarding the pollution level at  Delhi, India from Jan $1$, $2017$ - Jan $1$, $2018$. Here, we delve into the depth, and from empirical studies, it is evident that the behavioral patterns of societal perception can be associated with pollution level through the refinement of Spatio-temporal patterns, i.e., \textit{a. Time drift between social perception and ground truth and b. The sudden spurt of tweets without air pollution increment}. Refinement of such noise is required for the extraction of essential details from social sensors, which would effectively map the perceptions with the pollution level for a fine-grained pollution measurement. For solving such an issue, active perception monitoring can be performed by tracking influential users with the objective of getting a clear insight into the pollution level. Through monitoring a few influential users, the overall social perception can be inferred. Besides, specific trends might easily be comprehended by analyzing the contents of such users \cite{rabade2014survey}. There have been numerous methodologies \cite{rabade2014survey, riquelme2016measuring} for bringing out the factors which decide the influence of a user on other users on Twitter based upon information diffusion and user influence measurement. \\
Now, \textit{does the tracking of such influential users' pollution-related contents in Twitter results in a better correlation between social perception and the actual pollution level?} It is evident from past studies that the influential users set up the trends and regulates the community perceptions in Twitter regarding various facts like \textit{education, political issues, marketing, amongst others.} Hence, \textit{can their observations and knowledge assist in obtaining a better insight about the community perception towards pollution level?} These queries primarily motivated us to extract the influential users from collected pollution-related tweets of Delhi, India. Alongside, the perceptions of such influential users with the pollution level of CAAQM stations have been studied and illustrated.   
\vspace{-4pt}

\section{Contribution} \label{contrib}
It has already been illustrated in  Section \ref{intro} that the prime objective lies in relating the CAAQMS data with the tweets for obtaining a qualitative air quality monitoring system such that a tweet based framework can be utilized for predicting harmful levels of air quality where there are no CAAQMS. Therefore, for obtaining real-time trends regarding social perception towards varying air pollution levels, it is imperative to trace the most influential users who primarily regulate social behavior through his/her awareness. Detection of such users should be one of the foremost objectives for mining their knowledge or opinions for effective visualization of pollution status in urban or suburban regions. On the other hand, the utility of the perceptions of such influential users in context to the actual ground pollution level is also a matter of concern to train such a system. Considering such a key phenomenon, we have tried to extract the details of the  Twitter users who are influencing and persuading others to form their perception towards varying pollution levels. Besides, validation of such influential users' perceptions is also performed in order to study the effectiveness of such users' contents to the ground truth from CAAQM. In this regard, the contributions addressed by this current work are as follows:

\begin{enumerate}

\item We are developing an OSM (Twitter) based framework to understand the air quality in a developing region. This framework will further extend the coverage where there are no CAAQMS keeping \textit{Twitter as a "Sensor"}  
    
\item Considering the growing pollution at Delhi, India, we have collected pollution-related relevant tweets of the Delhi community using some hashtags which became popular during "Delhi Air pollution" from 2015 - 2018. 

\item Considering a particular snapshot of collected tweets of $2018$, the details of influential users are detected along with their influence score. A tailored TRank \cite{montangero2015trank} algorithm has been utilized for the extraction of such users based upon their \textit{Followers count, Re-tweeters' and Favorite Influence}. 

\item An exploratory analysis has been performed considering air quality from physical sensors at Delhi, India and the role of influential users to infer air quality from the tweets posted by them.

\end{enumerate}
\section{Data Acquisition: Social Perception \& Ground Truth} \label{dataacq}

Considering Delhi air pollution, there were some hashtags like \textit{airpollutiondelhi, delhismog, delhiairpollution, delhipollution}, which became popular on Twitter due to the massive rise in air pollution levels. The increase in air pollution in Delhi affects the people living in the specific urban and suburban regions, and there is a huge inflow of tweets related to these hashtags and becomes a trending topic within a short time interval. The first phase of the framework is to collect such pollution intent tweets from Twitter. We have used the python based library to access the Twitter API, Tweepy\footnote{http://docs.tweepy.org/en/v3.5.0/}, for such purpose. Each tweet collected using Tweepy contains several key-value pairs like id: Unique 64-bit integer tweet-id, created\_at: UTC time when the tweet is created, User\_id: Id of the user who created the tweet. Since the current work is targeted to monitor the pollution level of Delhi, only tweets relevant to  Delhi air pollution has only been considered. Besides, we have only considered the tweets posted in English. For the realization of such purposes, the API provides a filtering option that contains various fields like languages, keywords, locations based on which the unnecessary and irrelevant tweets were removed. As a result, the current work only considers English tweets related to Delhi. Furthermore, the filtration of tweets has also been conducted based upon a combination of keywords and popular hashtags like \textit{“NewDelhiairpollution”, “delhipollution” ,”delhismog” , “delhichokes”, “savedelhi”} . After defining all such parameters, the tweet-extraction has been completed. From the obtained the data set, it has been observed that around $166$K pollution intent tweets were posted from January 2015 through July of $2018$. Now, those obtained $166$K tweets are further analyzed based on the user type and their message content. On the other hand, we have gathered air pollution data from CPCB’s $33$ CAAQMs present at that time and the US Embassy station in Delhi from January $2015 -$ July $2018$. CPCB stations provide data on Principal Pollutants like $PM_{2.5}, PM_{10}, NO_{x}, SO_{2}$, and meteorological factors like Temperature, Humidity with a data granularity of 15 minutes. These ground truth data are also analyzed in a time frame ($2015 - 2018$) for monitoring the varying pollution levels in terms of $PM_{2.5}$ at Delhi, India. In the subsequent section, the methodology of finding influential users has been discussed in detail.       

\section{Finding Influential Users}
In this section, the methodology that has been used for the extraction of influential users from collected pollution-related tweets are discussed in detail. Here, the utility of three primary indicators of Twitter metric, i.e., \textit{a. Retweets, b. Favorites, and c. Followers} of a user have been explored in the data set to determine if he/she is an influential agent. The general psychology of human beings is to consider others' \textit{experience, opinion, and suggestions} before the creation of their perception towards any particular topic \cite{rabade2014survey}. Such phenomena are also realistic from the perspective of Twitter, where the users in the community are associated with a social relationship with others through \textit{follower-followee}. Among the users in the Twitter community, there are several experts and active tweeters who regulate the perceptions of the rest. Like all other fields, in context to air pollution monitoring, the detection of such users is also imperative for obtaining a better community perception through tracing the nodes actively contribution to understand the varying pollution levels. Several methodologies have been proposed in the past literature \cite{rabade2014survey,riquelme2016measuring,al2018analysis} for identification of influential users in Twitter. These methodologies are categorized as \textit{Centrality based measurement, Twitter API metrics-based, PageRank algorithm-based, PageRank with Twitter API, Topical sensitive, and Predictive measurement.} All such techniques utilize,
\begin{enumerate}
    \item Twitter API metrics like \textit{Followers, Retweets, Favorites, Mentions } or 
    \item The notion PageRank algorithm or
    \item Twitter Content Analysis or
    \item The combination of all such approaches for finding the user influence
    
\end{enumerate} Out of all such methodologies, a topic sensitive ranking approach, i.e., TRank \cite{montangero2015trank}, ranks users with three dimensions: \textit{follower relationship, retweets relationship, and favorites relationship}, has been considered. Unlike other techniques, TRank takes into account \textit{users, their tweets, and the simple attitude of human beings towards the usage of Twitter}. Hence, in the current work, we have used the notion of TRank using \textit{Followers' Influence of the user, Followers vs. Retweets of tweets posted by the user and Follower counts of Favourite users who liked the tweets of the user}. We have analyzed the collected $166$k tweets related to "Delhi Air Pollution" for realization users' influence in context to these three Twitter specific aspects. We have displayed the snapshot for the year 2018\\
\textbf{Retweet influence:} The original TRank stated that a user's influence is determined by the number of retweets per tweet a user has in his/her tweets, which is divided by the total number of followers a user has, thus if the ratio is greater than $1$. It implies that a user has influence beyond his follower region, which is a positive impact. However, this phenomenon has shortcomings. If we consider a user with one follower and single retweet with only one tweet, then he/she has retweet influence $1$ highest among all, which could lead to meaningless results. In order to compensate for this anomaly, we have introduced small customization to improvise the results. We have measured the \textit{activity (users' participation) and the retweet influence of a user}, we have analyzed each tweet of the user in the given time period and calculated the number of followers of every user who has retweeted a tweet of that user. The reason behind such an approach is to evaluate the followers of those active users who have read and found interest in the tweet and thus retweeted it. Hence, the retweet influence ($RI_i$) of a user $i$ can be obtained using the following equation. \\
\begin{equation}
    RI_i = \sum\limits_{j=1}^{n}\log(FR_{j})/ T_{t}(i)
\end{equation}
Here, $FR_{j}$ is the followers of the re-tweeters of $j^{th}$ tweet of user $i$ in a time interval $t$, $T_{t}(i)$ stands for total number of tweets posted by user $i$ within time interval $t$. Note, here $n$ stands for the total number of tweets the user $i$ posted within $t$. In current work, we have considered $t$ as the year $2018$. We have evaluated the influence of user $i$ based on the total number of tweets regarding air pollution (containing air pollution-related hashtags as discussed in the previous section) posted by that user within the time interval of $2018$. \\
\textbf{Favorite influence:} This is a major aspect for determining influence as "liking a tweet", which indicates that the tweet influences a user, and the count of the number of users who have liked a tweet is an important factor enhancing a user's influence. Hence, in order to calculate the favorite influence, we needed to fetch the followers of the users who have liked the tweet of the user $i$. We know that if a user likes a tweet, then it is available in his/her notifications which will be visible to all its followers.\\
\textbf{Followers influence:} This approach evaluates the number of followings of every follower of user $i$, as the number of active followings shall be inversely proportional to the follower influence considering human nature, there will be tweets from all the active followings at time interval $t$, so the probability that the user's tweet will be noticed is inversely proportional to the number of active followings. \\
Now, after the evaluation of \textit{Retweet, Favorite, and Follower} influence scores of each user $i$ within $t$, we extracted the \textit{mean influence score} of any user $i$ through considering the aforementioned influence measures.
\section{Results and Discussions: Insights \& Beyond} \label{discussion}
{

This section discusses the effect of customized TRank, using three indicators (i.e., \textit{a. Retweet, b. Favorite, and c. Followers influence}), which evaluates the combined influence score of any user. Detailed illustrations about such prominent users under these aspects have also been provided. Beyond that, an empirical study has been conducted for visualizing the correlation between a) CAAQM $PM_{2.5}$ related data - Tweets of Influential users obtained through customized TRank, b)  CAAQM $PM_{2.5}$ related data - Tweets of non-influential users obtained through customized TRank and c)  CAAQM $PM_{2.5}$ related data - Tweets of the users (both influential and non-influential). Consequently, the role of influential users in developing community perceptions towards the varying pollution level of Delhi, India, has also been elaborated.  
\begin{figure}[htbp]
\includegraphics[width=\linewidth]{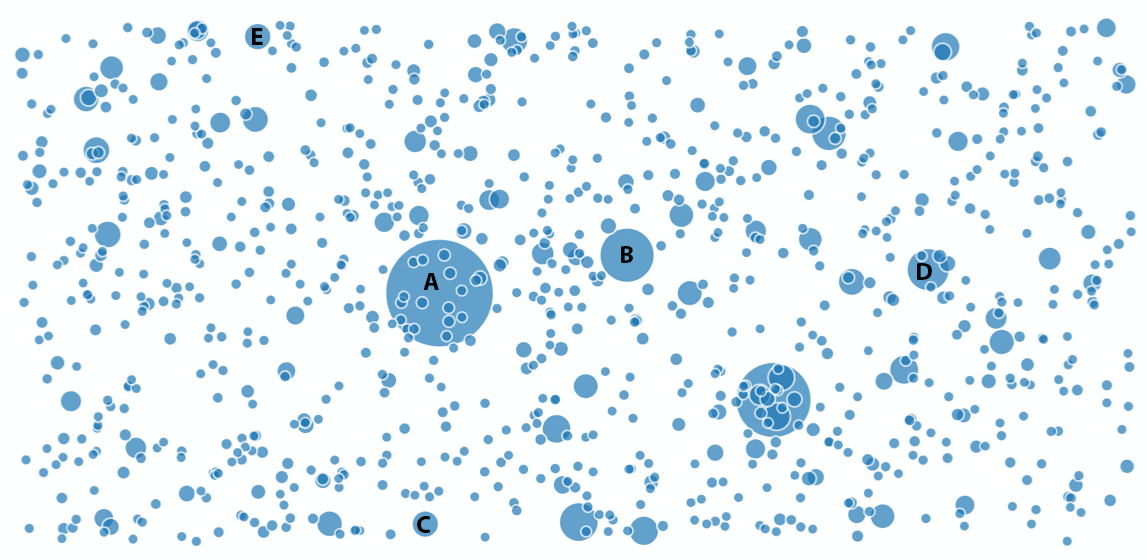}
\caption{Retweet Influence of Users Obtained for CPCB CAAQM $PM_{2.5}$ Data Distribution of Jan - July, 2018 for Delhi, India}\label{fig:awesome_image1}
\end{figure}

\subsection{Retweet Influence}

The graph in Figure \ref{fig:awesome_image1} is obtained by computing "Retweet Influence". The nodes represent the users. The nodes have \textit{user\_id and influence scores} as their attributes. The size of the nodes might vary, considering their influence scores. It can be observed that the biggest node \textit{(A, user-id = $711694309$)} has the highest influence score, which is $1333.951$ in terms of retweet influence. On the other hand, nodes like B (user-id = $2182353464$), C (user-id = $18548529$) ,D (user\_id = $2991581923$) and E (user-id = $33638320$) are having lower retweet influence scores $605.063$, $223.618$, $439.954$ and $225.321$ respectively.

\begin{figure}[htbp]
\includegraphics[width=\linewidth]{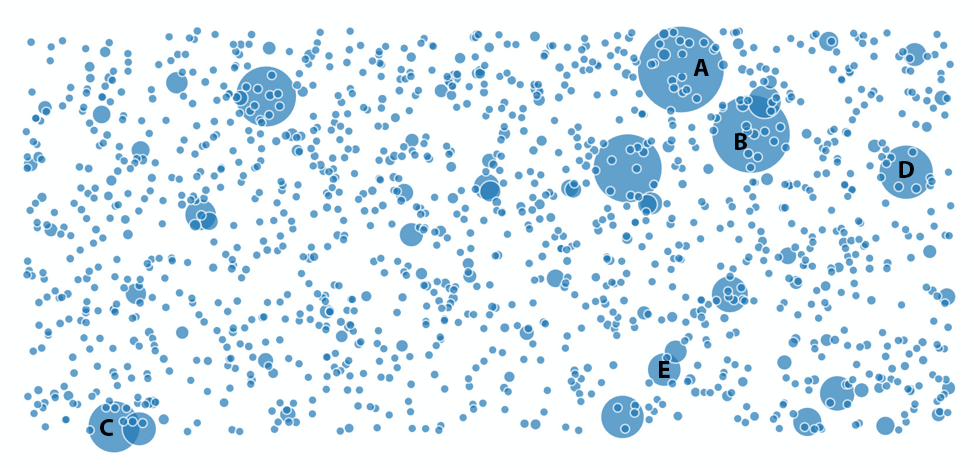}
\caption{Favorite Influence of Users Obtained for CPCB CAAQM $PM_{2.5}$ Data Distribution of Jan - July, 2018 for Delhi, India}\label{fig:awesome_image2}
\end{figure}

\begin{figure}[htbp]
\includegraphics[width=\linewidth]{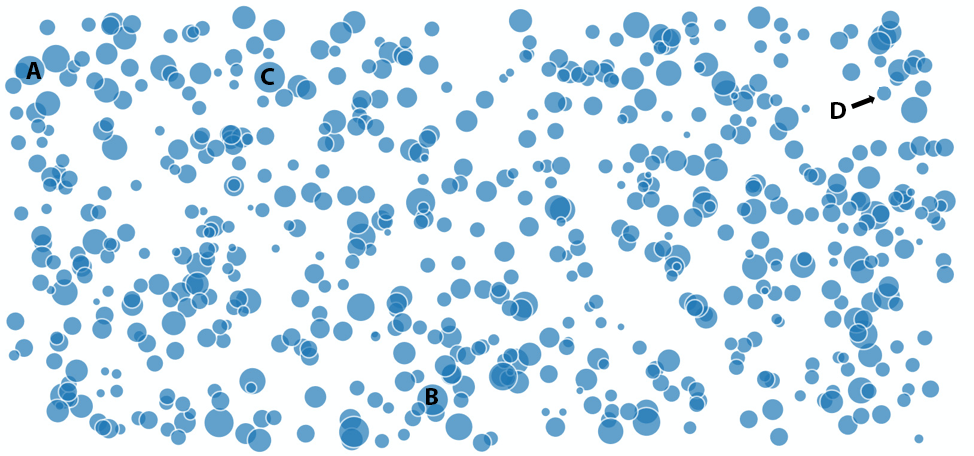}
\caption{Follower Influence of Users Obtained for CPCB CAAQM $PM_{2.5}$ Data Distribution of Jan - July, 2018 for Delhi, India}\label{fig:awesome_image3}
\end{figure} 
\begin{figure*}[htbp]
\includegraphics[width=\linewidth]{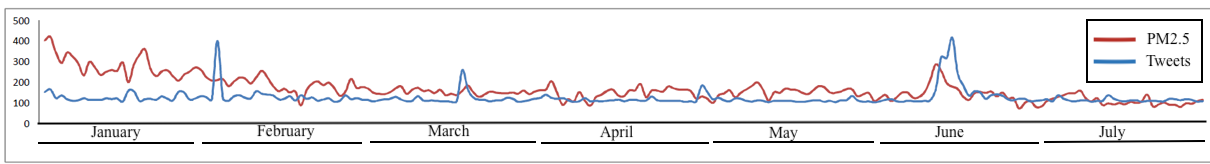}
\caption{Finding $PM_{2.5}$ level using the tweets of the All Users Obtained vs CPCB $PM_{2.5}$ Data Distribution of January - July, 2018 for Delhi, India}\label{fig:perceptionAll}
\end{figure*}

\begin{figure*}[htbp]
\includegraphics[width=\linewidth]{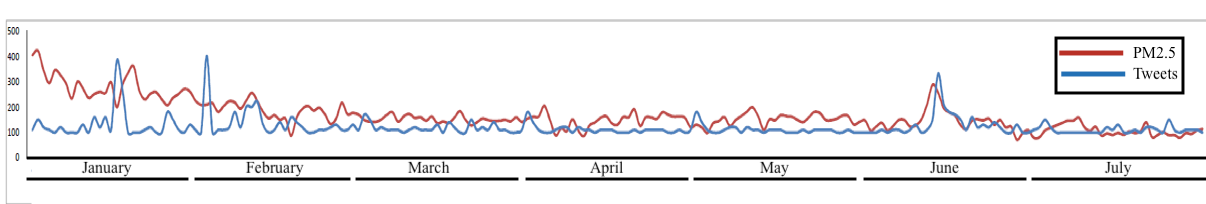}
\caption{Finding $PM_{2.5}$ level using the tweets of the Influential Obtained vs CPCB $PM_{2.5}$ Data Distribution of January - July, 2018 for Delhi, India}\label{fig:perceptionInf}
\end{figure*}

\begin{figure*}[htbp]
\includegraphics[width=\linewidth]{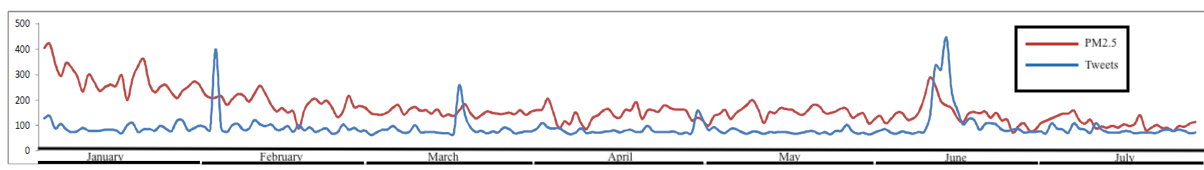}
\caption{Finding $PM_{2.5}$ level using only the tweets of the Non-Influential Users Obtained vs CPCB $PM_{2.5}$ Data Distribution of January - July, 2018 for Delhi, India}\label{fig:perceptionWithoutInf}
\end{figure*}

\subsection{Favorite Influence}
Figure \ref{fig:awesome_image2} is a visualization of the results obtained by computing "Favorite Influence". Here, it is evident that the biggest node (A, user id = $30417501$) has the highest Favorite Influence score of $1286$. The nodes B (user id = $711694309$) , C (user id = $318673863$) ,  D (user id = $741562726288351232$) and E (user id = $18548529$)  have influence scores  $1150, 719 ,754, 415$ respectively.

\subsection{Follower Influence}
The result obtained by computing "Follower Influence" has been depicted in Figure \ref{fig:awesome_image3}. Here, the biggest node (A, user-id = $134758540$) has the highest followers i.e. $11856069$. On the contrary, nodes B (user-id = $240649814$) , C (user-id = $39240673$) and  D (user-id = $155947001$) have $8940136 , 8341681, 1659$ followers respectively. Here, the size of the nodes in the direct implication of their follower count. Larger the node, more followers it has, and vice versa.

Please note, the placement of the nodes in Figure \ref{fig:awesome_image1}, \ref{fig:awesome_image2} and \ref{fig:awesome_image3} are completely random and have no significance.

\subsection{Correlation of User Perception \& Ground Truth}
After analyzing the three aspects \textit{Retweet Influence, Followers Influence and Favorite Influence}, and ranking all the users individually with respect to these three major aspects, we have gathered some interesting features when we compare a collective ranking of influential users combined with the CPCB data for Jan - July, $2018$. 

The plot in Figure \ref{fig:perceptionAll} shows the frequency of tweets for the months of 2018 (January-July) with respect to $PM_{2.5}$ levels. The tweets of all the users (non-influential users also) are considered for measuring the frequency.

From Figure \ref{fig:perceptionInf}, we can observe that the graphs of the tweets and the $PM_{2.5}$ level for each day have a close relationship between each other except for a few regions in the graph which deviate due to low granularity. This low granularity arises as the physical pollution sensors are far more fine-tuned than the calculated pollution level estimated from general perception of social media. We can understand when air pollution is high or low, but the fine-grained measurement level of physical sensors has not been achieved. However, we can see in the initial phase of pollution the level of tweets are not as high compared to the CPCB CAAQM $PM_{2.5}$, and they increase abruptly due to landmark reports from environmentalists, WHO, CPCB, NGO’s and news agencies which lead to extensive amount of fact-based tweets. It had also been reported previously in \cite{pramanik2018aircalypse} and needs to be qualitatively dealt with. There also lies a region in which the frequency of tweets is abruptly increasing and decreasing due to deviation from the topic of discussion, and unrelated topics emerge as a new trending topic. Additionally, in some instances, the frequency of tweets becomes abruptly high after a period of high pollution because it takes time to form a social perception when compared to air-pollution sensors used to measure pollution for each day. \\
In Figure \ref{fig:perceptionWithoutInf}, the plot has been formed between the non-influential users and the frequency of tweets for each day. There is no whatsoever tightly bounded relation between the level tweets and $PM_{2.5}$ concentration. Sometimes the frequency of tweets is considerably low, and in irregular occurrences, there are some non-fixed steep rises that are difficult to be judged and rationalize. There is only one fixed occasion where the $PM_{2.5}$ concentration has shown a similar behavior with respect to the frequency of tweets; the rest are pretty much distorted. In Figure \ref{fig:perceptionAll}, the two curves $PM_{2.5}$ and tweet frequency appear to be closer as compared to the results obtained in Figure \ref{fig:perceptionWithoutInf}, where influential users are excluded.

Hence, considering the influential users plays a critical role in the development of the framework as it improves the accuracy of the perception in online social media i.e., Twitter and thus improving its performance \textit{as a sensor}.
}

\section{Comments}{

This is work in progress, where we want to create the end to end framework of using social media like Twitter as an air quality sensor. The issues that still need to be dealt with includes:
\begin{enumerate}
    \item Geo-location of each tweet: Only 3-5\% of the overall tweets are geo-tagged. Hence, it is not easy to locate the origin of the tweet to the street level in a city. Developing a geo-location-based tweet framework will help in fine-grained estimation of air quality. 
    \item Considering the importance of general users rather than the retweet importance of a celebrity or news medium accounts. This is because many users follow the celebrity and news medium accounts; hence it easily gets more retweets and favorites. The general users need to have a normalization when compare to the celebrity and news accounts.
\end{enumerate}

}
\bibliographystyle{unsrt}  
\bibliography{references}



\end{document}